%% file: article.tex
\documentclass[a4paper,titlepage]{article}

\usepackage{arxiv}
\usepackage[utf8]{inputenc} 
\usepackage[T1]{fontenc}    
\usepackage{hyperref}       
\usepackage{url}            
\usepackage{booktabs}       
\usepackage{amsfonts}       
\usepackage{nicefrac}       
\usepackage{microtype}      
\usepackage{listings}
\usepackage{graphicx}
\usepackage{subcaption}
\usepackage{color}
\usepackage[svgnames]{xcolor}
\usepackage[title]{appendix}

\newcommand*{\fullref}[1]{\hyperref[{#1}]{\mbox{\autoref*{#1}~}``\nameref*{#1}''}}
\newcommand{\code}[1]{\mbox{\lstinline{#1}}}
\title{Loop Programming Practices that Simplify Quicksort Implementations}

\author{Shoupu Wan \\
\texttt{wanshoupu@gmail.com} \\
}

\begin{document}
    \maketitle

    \begin{abstract}
        Quicksort algorithm with Hoare's partition scheme is traditionally implemented with
        nested loops.
        In this article, we present loop programming and refactoring techniques
        that lead to simplified implementation for Hoare's quicksort algorithm consisting of a
        single loop.
        We believe that the techniques are beneficial for general programming and may be used for
        the discovery of more novel algorithms.
    \end{abstract}

    \keywords{Quicksort \and Loop rotation \and Nested loop thinning \and Cascading conditional
    \and Sentinel}

    \section{Introduction}\label{sec:introduction}
    Loops are one of the most widely used programming constructs featured
    in almost all programming languages.
    A loop is an productivity amplifier.
    With nominal overheads (\textit{e.g.}, state-registering variables, \textit{etc.}),
    the static body of a loop can be reused for unlimited number of times.

    A key building block as loop, one would suppose its best practices should have been widely
    known.
    On the opposite, however, the best practices for loop has been so largely ignored that
    haphazardly constructed loops with duplication issues is not uncommon even in production code.
    Common problems in loop programming include, but are not limited to, duplicate code, nested
    loops, leaky loop variables, and oversized initialization.
    I will explain each of them next.

    Duplicate code here refers to duplication between code in the loop body and code before (after)
    the loop.
    It is the biggest problem in loop programming and is the most common root causes for bugs.
    Duplicate code anywhere is bad.
    But duplicate code of this type is harder to realize or get rid of.

    Uses of nested loops are sometimes controversial.
    Many readers are ready to argue about this.
    Anyhow, what is wrong with nested loop?
    For almost all cases, nested loops bring complexity rather than convenience,
    obstruct readability rather than facilitating it.
    It turns what could have been coded as different components into monolithic mess and
    discourages code reusing.
    deepen the coupling rather than reducing it and discourage code reusability rather than encouraging it.

    Loop variables refers to variables used to register loop state.
    Many developers rely on exposed variable(s) to implicitly pass information from loop to
    subsequent program.
    However, uncontrolled exposure of loop variable to subsequent code is a violation of the
    encapsulation principle.
    While sometimes convenient, it usually does more harm than good.

    Loop initialization is the prelude code needed for instantiating initial loop state.
    It should be small and light-weight.
    rather than light-weight, succinct ones.
    However unseasoned programmers may code disproportionally heavy initializations.
    With due skills and perseverance, loop initializations can be made succinct one-liner.

    It is beyond the scope of this article to address all the topics.
    Instead we will focus on duplicate code and nested loops.
    We will present two pillar loop programming techniques---`loop rotation' and `Nested loop
    thinning' which I found are effective in fighting against above programming foes.
    These are the two pillars for loop programming.
    Proper use of them help developers avoid commonly made mistakes.

    We have tried these techniques on several case studies.
    Particularly, we will apply them to simplify traditional quicksort algorithm to prove the
    effectiveness of these techniques---the implementation of quicksort algorithms.
    First invented in 1960, quicksort has been studied and analyzed well over the
    years\cite{Hoare1961,Hoare1962,sedgewick:1978,Bentley:1986:PP}.
    As one of the earliest `divide-n-conquer` algorithms, quicksort has become the
    \textit{de facto} sorting algorithm in practice for its excellent expected performance.
    Recently named one of the top $10$ algorithms of $20$th century, quicksort has a
    profound influence on the history of programming\cite{top10algorithm:2000}.

    We will walk you through the multi-stage refactor process that leads to the discovery of a
    brand-new implementation of Hoare's quicksort algorithm.
    We use Python as the working language in this article.

    \section{Loop Rotation}\label{sec:loop-rotation}
    \input{loop-rotation}

    \section{Nested Loop thinning Technique}\label{sec:loop-thinning-technique}
    \input{nested-loop-thinning-technique}

    \section{Case Study: quicksort}\label{sec:nested-loop-quick-sort}
    \input{case-study-quicksort}

    \section{Experiment}\label{sec:experiment}
    \include{quicksort_perf_cmp}

    \section{Conclusion}\label{sec:conclusion}
    In this article, we presented a couple of techniques for optimizing loop constructs in
    high-level programming languages.
    Taking advantage of the circular symmetry of loop constructs, loop rotation may be applied to a
    loop to either reduce code duplication (forward rotation) or to shift certain part of the
    code within the loop body (reverse rotation).
    Another technique is loop thinning for simplifying nested loop complications.
    These are two of the empirical techniques that helps programmers achieve software development
    best practices.
    As an example, we applied these techniques in simplifying a traditional implementation of
    Hoare's quicksort algorithm.
    We provided the simplified implementation in C++.
    More generally, the programming techniques we developed in this article are applicable to all
    programming languages that supports loop and conditional constructs.
    \begin{appendices}
        \appendix
        \include{partition-functions}
        \include{use-sentinel}
    \end{appendices}
    \bibliographystyle{unsrt}
    \bibliography{article}
\end{document}

%% file: loop-rotation.tex
\begin{figure}[htb]
    \centering
    \begin{subfigure}{.3\textwidth}
        \includegraphics[width=\textwidth]{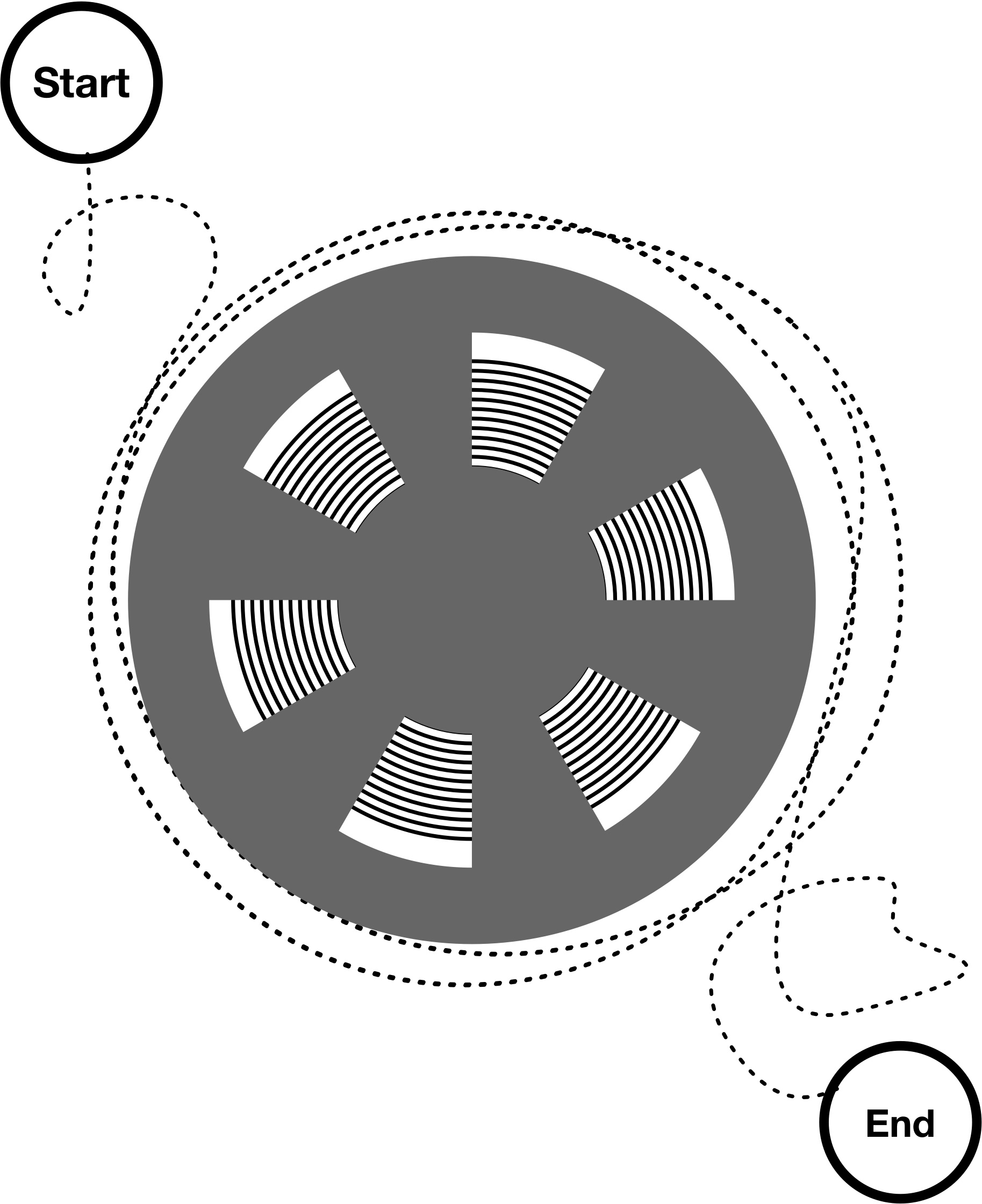}
        \caption{\label{fig:loose-spool}}
    \end{subfigure}
    \begin{subfigure}{.3\textwidth}
        \includegraphics[width=\textwidth]{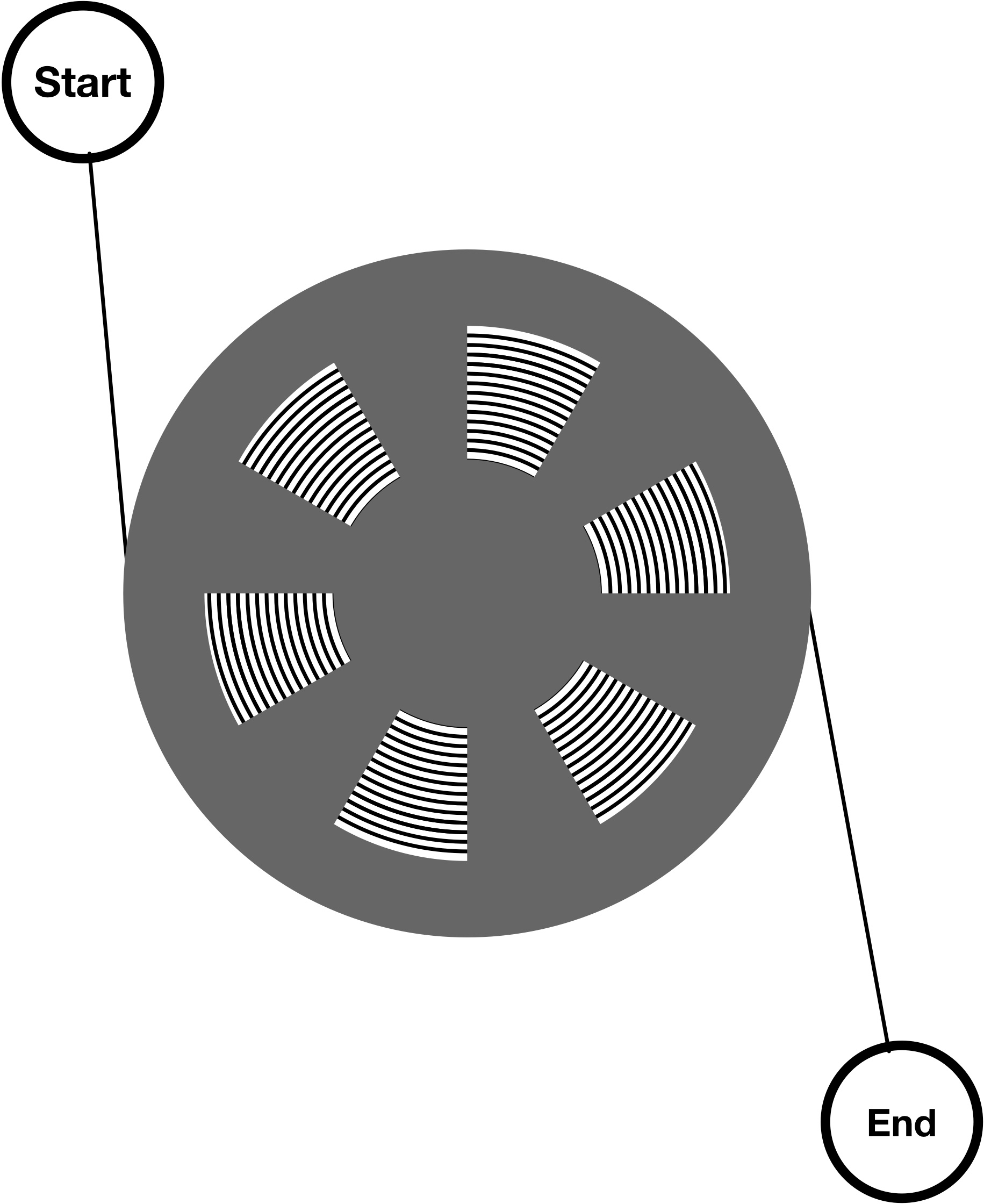}
        \caption{\label{fig:tight-spool}}
    \end{subfigure}
    \caption{\label{fig:spool-loop} Using spool as a metaphor for loop construct in program,
    duplicate code around a loop would be like tightening messy yarns into neat spool.
    `Loop rotation' can be pictorially depicted as winding up loose yarns onto a spool.
    \protect\subref{fig:loose-spool}: Duplicate, sloppy code is analogous to loose, unorganized
    yarn around a spool.
    \protect\subref{fig:tight-spool}: A neat loop is analogous to a tidy spool.}
\end{figure}

Coming with the productivity amplification power is the coding complexity of loops.
There are two key observations about loops.
First, a loop seldom lives in vacuum.
Loops are often used as an embedded subunits in a program just like an organelle in a living cell.
So it has to get along with its neighbors.
As a result, a large part of loop programming is to ``fit it in''.
Secondly, there are many moving parts in a loop construct and they are tightly coupled, changes in
one necessarily demand changes in others.
As we discussed in ~\autoref{sec:introduction}, the biggest problem with loop programming is
duplicate code in and around the loop.
The foremost goal in implementing loop is to reduce duplicate code.
`Loop rotation' is an important technique for that purpose.

It is beneficial to visualize a loop construct as a spool.
Loops are to program as spools are to yarn---just as spools are used to organize yarn, loops are
invented to organize program.
Behind this analogy is an important symmetry with respect to circular shifts---the rotational
symmetry shared by them.
Aside from the point of entrance and one or more exits, a loop or a spool may be mathematically
represented by a circular sequence, that is, a sequence with rotational symmetry.
One can thus take this rotational degrees of freedom to one's advantage to decide where
to enter and exit the loop.
Coalescing duplicate code in a loop is analogous to
winding up loose and messy yarn into a tidy spool (~\autoref{fig:spool-loop}).

Let's take a board game as an example to illustrate loop rotation.
Imagine that the following subroutines are ready to use:
\code{Init}: Initialize game;
\code{BD}: Draw board;
\code{PR}: Print legends and prompts;
\code{UI}: Take user inputs;
\code{EX}: Execute user inputs;
\code{CM}: Compute moves;
\code{PO}: Poll game status;
\code{End}: End of game.
~\autoref{board_game_dupcode} shows pseudo-code for the driver program.
The duplicate code between the loop and in the vicinity of the loop is conspicuous.
With a ``loop rotation'' procedure (to be explained below), the code may be refactored into
~\autoref{board_game_rotated}.

\begin{minipage}[t]{.4\textwidth}
\begin{lstlisting}[language=python,
    caption={\label{board_game_dupcode}Board game driver program}]
Init
BD
PR
UI
EX
CM
while PO:
    BD
    PR
    UI
    EX
    CM
End
\end{lstlisting}
\end{minipage}\hfill
\begin{minipage}[t]{.4\textwidth}
\begin{lstlisting}[language=python,
    caption={\label{board_game_rotated}Refactored driver program}]
Init
while True:
    BD
    PR
    UI
    EX
    CM
    if PO:
        break
End
\end{lstlisting}
\end{minipage}

\begin{figure}[htb]
    \centering
    \begin{subfigure}{.3\textwidth}
        \includegraphics[width=\textwidth]{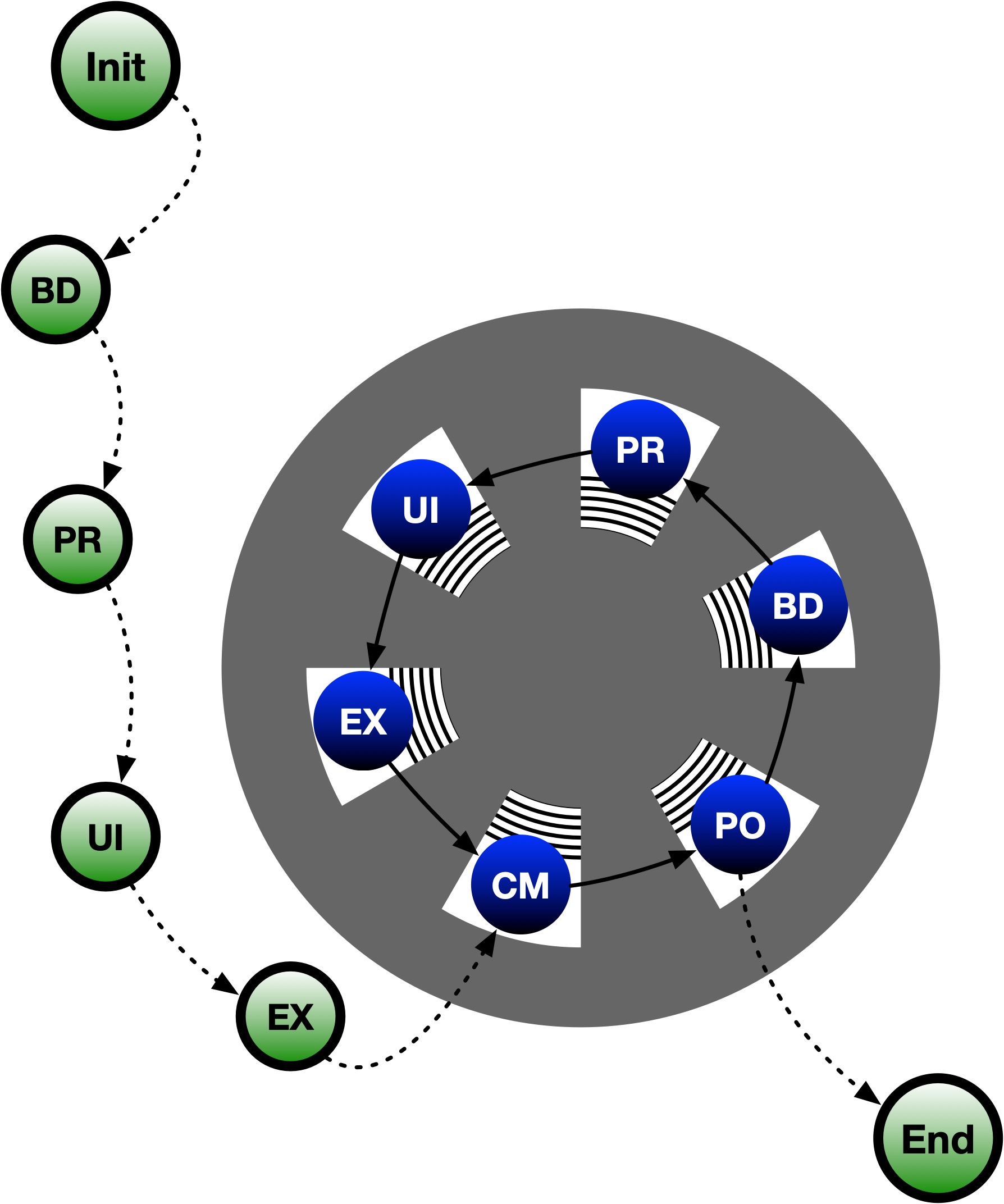}
        \caption{\label{fig:ferris-wheel-dup-node}}
    \end{subfigure}
    \begin{subfigure}{.3\textwidth}
        \includegraphics[width=\textwidth]{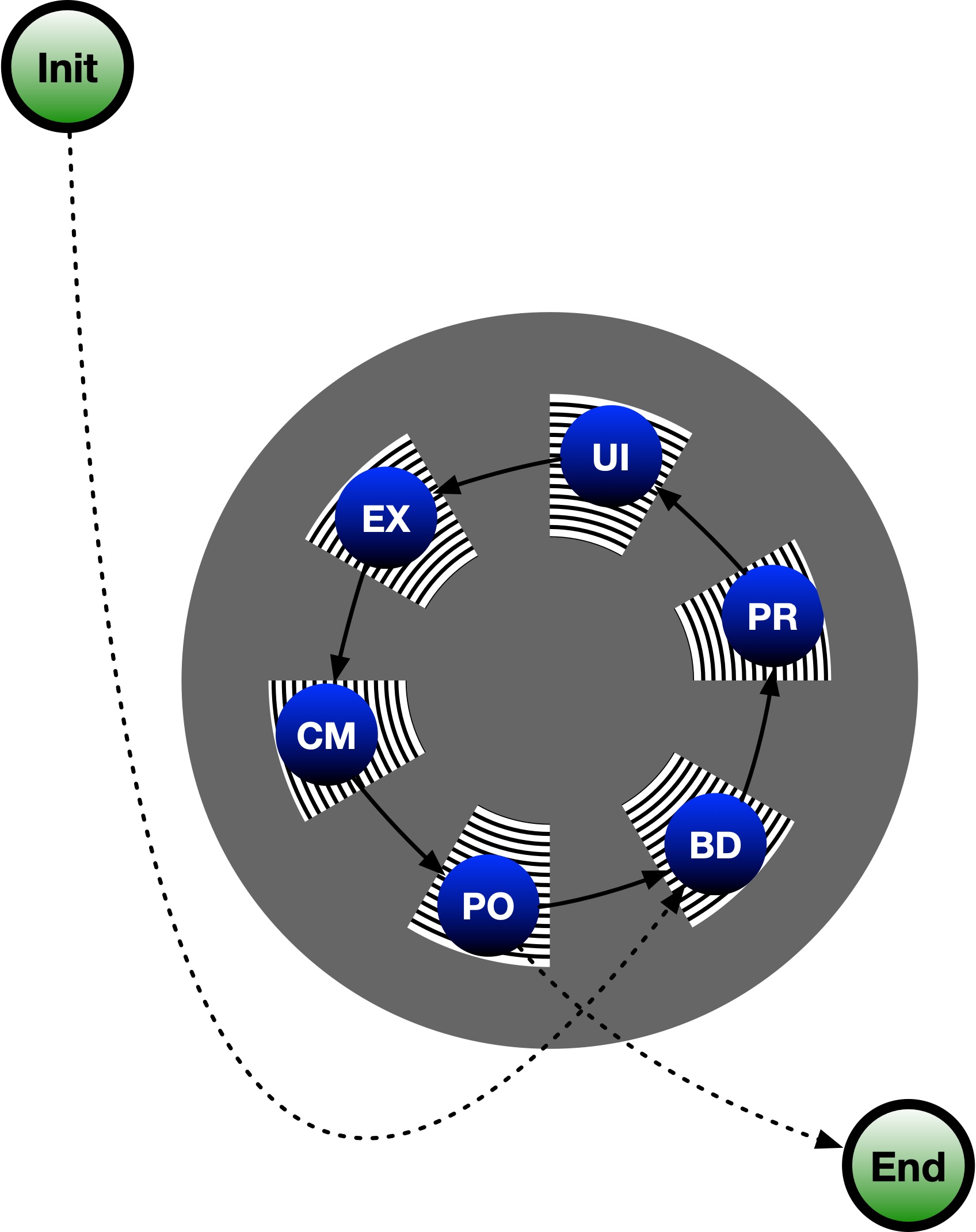}
        \caption{\label{fig:ferris-wheel-rotated}}
    \end{subfigure}
    \caption{\label{fig:loop-rotation}
    Pictorial example of how loop rotation coalesces duplicate code.
    Legend: blue are loop statements, green are non-loop statements.
    The statements ``\code{BO}'', ``\code{PR}'', ``\code{UI}'', and ``\code{EX}''
    are duplicated between pre-loop and loop.
    They may be coalesced by a loop rotation.
    \protect\subref{fig:ferris-wheel-dup-node}: Original program;
    \protect\subref{fig:ferris-wheel-rotated}: Rotated program.}
\end{figure}

Looking at ~\autoref{fig:loop-rotation}, as we rotate the loop, the duplicate code can be aligned
and coalesced line by line.
As can been seen in ~\autoref{fig:ferris-wheel-dup-node},
the last statement in the loop and the last statement before the loop
are verbatim duplicate and aligned.
As such, we `roll up' the spool so that the two can be coalesced.
This process can be repeated until all the duplicates are coalesced
(~\autoref{fig:ferris-wheel-rotated}).

Getting rid of duplicate code is one of the main application scenario for ``loop rotation''.
Another application scenario for the ``loop rotation'' procedure is for the effect of
shifting code within a loop, for example, to move a portion of the code from the beginning to the
end.
This is often called ``reverse rotation'' of the loop.
As a side effect, doing so will result in duplicate code.
Given that this prepares ways for subsequent refactors, this adverse effect is often paid
off with larger optimizations.
In practice, one may start with the more malleable ``\code{while (true)}'' or
``\code{for (; ;)}'' loops which
helps one focus on getting a correct program first.
Once \emph{a} working and flexible code has been secured,
the `loop rotation' technique can be used to fine-tune the program.

%% file: nested-loop-thinning-technique.tex
Every developer writes nested loops now and then.
Many times, nested loops appear compelling and inevitable.
While they may solve our problem, nested loops inflict on a program
unnecessary complexity, obstruct code readability, and bring in `soft duplication'.
The presence of nested loops also thwart optimization at the compiler level.
For these reasons, explicit use of loops at high level programming should be avoided
altogether\cite{data-engineering-avoid-loop}.
Moreover, getting rid of nested loops itself may not seem significant improvement.
The optimizations that are made accessible after getting rid of nested loops dwarfs the improvement
brought about by getting rid of nested loops itself.
Like `Candy Crush Saga' game at certain critical point, a single move may unlock an
avalanche of advantageous moves.

\begin{figure}
    \centering
    \includegraphics[width=0.4\textwidth]{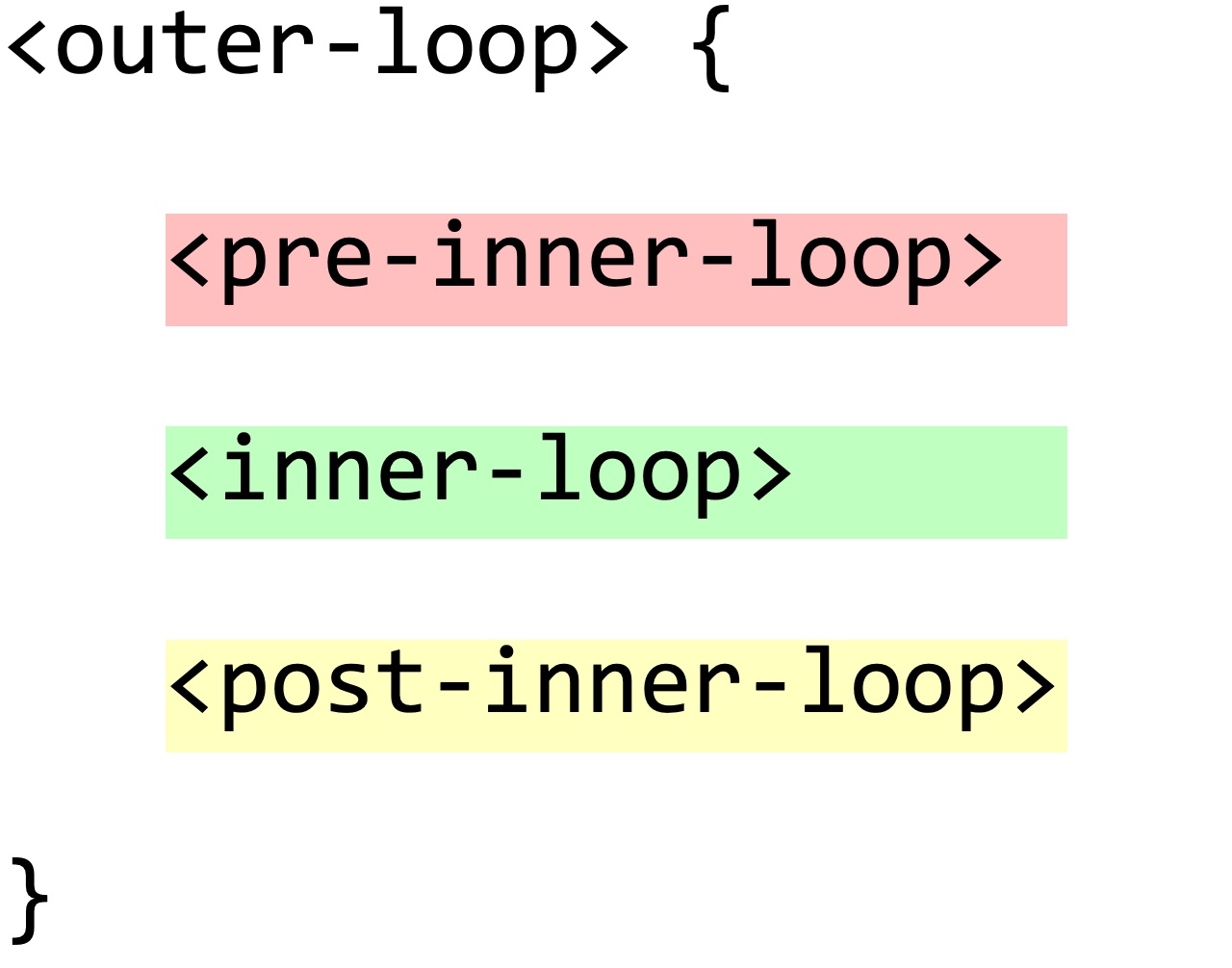}
    \caption{\label{fig:segment-for-loop-thinning}
    Sectioning of the body of nested loops:
    \emph{pre-inner-loop} statements, the inner loop itself, and the \emph{post-inner-loop}
    statements.
    Each section receives different treatment during the `nested loop thinning' process.}
\end{figure}

In this section, we are to present a process that coalesces nested loops into single-layered loops,
which we will name as `Nested loop thinning technique'.
For sake of argument, the body of the outer loop is divided into three
sections---\emph{pre-inner-loop} section, \emph{inner-loop} section, and
\emph{post-inner-loop} section---depending on their relative position
in respect to the inner loop (as shown in ~\autoref{fig:segment-for-loop-thinning}).
Cramming the functionalities of nested loops into a single loop is no easy task.
As a price, the process almost always results in one or more of the following:
\begin{itemize}
    \item extra conditional constructs
    \item extra tests added to existing conditional constructs
    \item more dynamic loop pacing
    \item more boundary-condition-handling logic
    \item auxiliary data structure, such as queue or stack
\end{itemize}

The gist for nested loop thinning process is as follows:
\begin{enumerate}
    \item Preparation stage
    \item Loop rotation to shift around components in the loop body
    \item Reconstruction of loop body using conditionals
\end{enumerate}
But we will go through them one by one.
One will see this pattern again and again for nested loop thinning in practice.

First, some preparatory measures may be taken beforehand
to reduce the friction during the refactor process.
Compound statement is a good example to be dealt with in this step.
Compound statements may get in your way of refactoring for multiple reasons.
The most obvious one is that a compound statement
need to be broken up and sent to different places after refactoring.
State-changing, non-idempotent conditional expressions, such as \code{if (--i < 0)},
are even more lethal because each evaluation of the condition ratchets the state of the loop.
For example,

\begin{minipage}[c]{.3\textwidth}
    \begin{lstlisting}[language=C]
        while (++i < LIMIT);
    \end{lstlisting}
\end{minipage}
\hfill
\begin{minipage}[c]{.3\textwidth}
    shall be expanded to
\end{minipage}
\begin{minipage}[c]{.3\textwidth}
\begin{lstlisting}[language=C]
do {
    ++i;
} while (i < LIMIT);
\end{lstlisting}
\end{minipage}

before the start of loop thinning refactor.
Other types of compound statements, such as ``\code{if v := a[i]; v < LIMIT}'' in Golang, shall be
preprocessed similarly.
Because of the structural similarity between \code{while}-loops and conditionals.
\code{while}-loop readily lends itself to ``nested loop thinning'' process.
For this reason, \code{for}-loop are often converted to \code{while}-loop during preprocessing.

In the second stage, one is to move pre-loop statements, if any, out of the way.
To that end, reverse loop rotation may be used
to unwind the pre-loop statements (see ~\autoref{sec:loop-rotation}).
The second step depends on the inner loop construct.

Finally, it the reconstruction of the loop body using conditional rather than nested loops.
If the inner loop is an unconditional loop as `\code{while True}',
\code{break} statements (or similar) are almost always present and most likely
in a conditional statement somewhere in the loop body unless
it is intended to be a non-typical loop construct.
One should replace the \code{break} statement with the post-inner-loop statements.
The inner loop can then be stripped away.
Otherwise, if the inner loop comes with a non-trivial termination condition,
then the inner loop can be converted to a conditional directly,
\code{while} $\rightarrow$ \code{if},
while the post-inner-loop statements are wrapped away in an \code{else}-clause of it.

Regardless of the venue taken, new conditional statements are inevitably formed or extended
and `cascading conditional construct' are the best way to organize them.
`Cascading conditional construct' consists of ordered sequence of exclusive conditional
statements such as ``\code{if .. elif .. else}''.
For more information, please refer to relevant chapters in Reference ~\cite{Wan:book}.
Throughout the process, one shall pay special attention to execution-path-shunting statements,
such as \code{break} and \code{continue}, if any.

After `nested loop thinning', some cleanup may be performed to comply with convention, code
style, or just for cosmetic reasons.

One disclaimer is that the `nested loop thinning' process does not always prevail.
There are cases where the process is not applicable.
Certain criteria must be met for `nested loop thinning' to be applicable.
First, there must not be intermediate layers, such as
conditional, between the inner and outer loops.
Secondly, execution-path-shunting statements, such as \code{break}, cannot be present in the
pre-inner-loop section.
In what follows, we are going to demonstrate application of `nested loop thinning' technique
on the quicksort algorithms.

%% file: case-study-quicksort.tex
Quicksort is one of the pivotal sorting algorithms widely used by modern software.
The Hoare's scheme was the first partition scheme that came
with the original invention of this algorithm\cite{Hoare1961,Hoare1962,sedgewick:1978}.
Traditionally, Hoare's partition scheme has been implemented with nested loops.
Later one, Lomuto's partitioning scheme was invented whose implementation is much simpler with only
one loop\cite{Bentley:1986:PP}.
However for certain edge cases, Lomuto's quicksort algorithm does not perform well.
It is natural to ask if one can implement the Hoare's partition scheme with the simplicity of or
close to the Lomuto's.
Armed with the loop programming techniques presented in this article, let us give it a try.
First let us lay the foundation of the implementations of the quicksort algorithm.

\subsection{Recursive implementation of quicksort}\label{subsec:recursive-impl-quicksort}
At the high level, a recursive quicksort implementation may be as follows
\begin{lstlisting}[label={lst:QuickSortOutline},language=C,
caption={An implementation of quicksort algorithm.}]
void qsort(int* s, int* e) {
     if (e - s < 2) return;
     int* p = part(s, e);
     qsort(s, p);
     qsort(p + 1, e);
}
\end{lstlisting}
where arguments \code{s} and \code{e} are the starting and ending pointers to the input array.
This function invokes `\code{int* part(int* s, int* e)}' which is a function stub for array
partition which will be discussed in detail below.
For single-element arrays, quicksort function is no-op which is a base case.
For na\"{i}ve implementations, this the only base case and it is sufficient.
But for more sophisticated implementations or to reduce partition overhead, more base
cases are used to address under-sized arrays (\textit{e.g.}, arrays of size $\leq 5$ but $\geq 1$).

In plain English, this is how quicksort algorithm works:
\begin{quote}
    If the array contains fewer than $2$ elements (the base case), return as is.
    Otherwise, invoke partition function to partition the array into two subarrays and an
    element (the \emph{pivot}).
    Each subarray is subject to the quicksort function again so on and so forth
    until they are all reduced to the base case.
    At the return of the function, the entire array is sorted.
\end{quote}
The implementation of the partition function is key to the quicksort algorithm.
The partition function does three things:
\begin{enumerate}
    \item Pick a pivot from among the array and set it aside;
    \item Use pivot as a benchmark, partition the rest of the array into two subarrays
    with smaller (or equal) ones on the left and greater (or equal) ones on the right;
    \item Put the pivot element back in between the subarrays.
\end{enumerate}

The subarrays resulted from partition may not be equal sized which is called partition skewness.
Partition skewness has an adverse impact on the performance of quicksort algorithm.
In all practical implementations of the partition function for quicksort, some type of pivot
selection strategy is needed to prevent partition skewness.
Common and proven practices are random selection or ``median-of-three''
technique\cite{sedgewick:1978,sedgewick1977}.
Of course, to apply the ``median-of-three'' technique, the array must exceed a minimum size.

With this said, we are ready to discuss implementation of quicksort and its partition function.
Admittedly, implementing quicksort is quite tricky.
Among the quicksort partition schemes, best known are
Hoare's scheme and Lomuto's scheme\cite{Hoare1961,Bentley:1986:PP}.
The main difference between them lies in how the array is traversed.
In Hoare's scheme, two pointers, one from each end of the array, step toward each other;
whereas in Lomuto's, two pointers, each on its own pace, start off the left end of the array
and step rightward.

Among the quicksort algorithms, there are two main variants---those
by Tony Hoare and by Nico Lomuto, respectively.
Hoare's quicksort scheme has robust and optimal performance but its
implementation has been quite involved.
Lomuto's scheme, on the other hand, is straightforward to implement and easier to follow.
However its performance may degrade catastrophically for certain edge cases.
Comparing the two,
one cannot help but wonder if there is an implementation
that is as robust and performant as the traditional implementation of Hoare's quicksort algorithm
and at the same time as succinct as that of Lomuto's.
That is going to be the focus of rest of this article.

\subsection{Implementation of Lomuto's Partition Scheme}\label{subsec:impl-of-Lomuto-part-fun}

\begin{lstlisting}[label={lst:singleLoopLomuto},language=C,
caption={Implementation of Lomuto's partition scheme for quicksort.}]
int* part(int* s, int*e) {
    int * p = s;
    for(int* i = s; i < e; i++ ) {
        if (*i < *s) {
            swap(++p, i); //swap the out-of-place elements
        }
    }
    swap(s, p);
    return p;
}
\end{lstlisting}
Let's start with the relatively simpler Lomuto's partition scheme.
In Lomuto's partition scheme, the two pointers have distinct tasks.
The one running in front, variable \code{i}, is responsible to discover out-of-place elements.
The one behind, \code{p}, guards the partition boundary.
When \code{i} discovers an out-of-place element, \code{p} makes
room and places it by a swap and the partition process continues.
The code is shown in ~\autoref{lst:singleLoopLomuto}.
Once one understands the code, implementation becomes highly consistent and intuitive.
One seldom fails implementing even for a customized applications.

Notably, implementing this partition scheme only needs one loop.
However the simplicity is no free lunch.
In fact, for certain edge cases, Lomuto's partition scheme suffers severe performance penalty,
\texttt{e.g.}, arrays with a large number of identical elements,
in which the Lomuto's quicksort algorithm degrades close to quadratic runtime.
The root cause leading to this degradation lies in the asymmetric traversal of the array which
inevitably leads to partition skewness.
With the Lomuto's quicksort partition function, let's come back and study the more sophisticated
implementation---the symmetric Hoare's partition scheme.

\subsection{Implementation of Hoare's Partition Scheme}\label{subsec:impl-of-qsort-part-fun}

Invented along with the quicksort algorithm, the Hoare's partition scheme predates Lomuto's
historically\cite{Hoare1961,Hoare1962,sedgewick:1978}.
Because of its symmetric traversal, Hoare's partition scheme successfully avoids the drawback of
Lomuto's.

Visually speaking, Hoare's partition scheme employs two pointers, \code{s} and \code{e}, starting
off the opposite ends of the array, push through the array toward each other,
given that a pivot element has been placed at the beginning of the array.
As in Lomuto's partition function, these pointers also stop at `out-of-place' elements.
When both stop, the `out-of-place' elements are swapped to where they belong.
Then the pointers are on their way again so on and so forth until they meet or cross.
At last the pivot element is swapped into its final position and a pointer to this final position
is returned.

While it appears a minor change from Lomuto's scheme, the Hoare's partition comes with immense
implementation complexity.
So much so that for a long time how Hoare's partition scheme work
remained an enigma\cite{Bentley-Beautiful-code}.
There are so many changing variables and so much coupling among them that once in a
while, each attempt of implementing it may end up with a different solution.
Even worse than that, when something goes wrong, one is often clueless as to what is wrong.
Also, it is extremely hard, if not impossible, to devise a test case to hit an elusive bug.

But in stark contrast to the numerous slightly differing implementations, all known implementations
have so far unanimously used $3$ nested loops: one outer loop and two sequential inner loops.
This feature is so commonplace that it has become the stereotype of quicksort.


With the Hoare's partition scheme, a commonly found implementation for the partition function is as
follows.
\begin{lstlisting}[label={lst:nestedLoopHoareMethod},language=C,
caption={Implementation of Hoare's partition scheme for quicksort.}]
int* part(int* s, int*e) {
    int* const pivot = s;
    while(true) {
        while(s < e && *++s < *pivot);  // find the next element >= pivot
        while(pivot < e && *--e > *pivot); // find the prev element <= pivot
        if (s < e) {
            swap(s, e); //swap the out-of-place elements
        } else {
            swap(pivot, e);
            return e;
        }
    }
}
\end{lstlisting}
While we have given an outline of the working of Hoare's partition, many choices remain to be
made in regard of ``how, when, and what''.
As such, pitfalls lay in wait every now and then throughout the implementation process.
We will leave the discussions of the implementation process of ~\autoref{lst:nestedLoopHoareMethod}
encountered during this implementation in ~\autoref{sec:impl-varieties}.

\subsection{Thinning of Nested Loops}\label{subsec:unwind-hoare-part}

Now we are going to use the techniques presented earlier to transform the traditional
implementation of Hoare's partition scheme and get rid of the nested loops.

The immediate difficulty is how to take apart the densely packed conditional construct:
\begin{lstlisting}
    while(s < e && *++s < *pivot);
    while(pivot < e && *--e > *pivot);
\end{lstlisting}
The loop conditions here are awkwardly complicated and make the 'nested loop thinning'
(outlined in ~\autoref{sec:loop-thinning-technique}) nontrivial.
The main difficulty lies in the dilemma---when condition suits, we need to switch to
alternative execute path;
but when we do, the pre-incremental statements would have ratcheted
the state variables `(\code{s}, \code{e})' one step too far.

Measure must be taken to break up the pre-incremental statements before we can proceed any further.
We follow a two-step conversion procedure: first unfold to `\code{do-while}' and then,
in turn, rotate to `\code{while}' as shown in ~\autoref{quicksort_inner_loop_do_while} and
~\autoref{quicksort_inner_loop_while}.

\begin{minipage}[t]{.4\textwidth}
\begin{lstlisting}[caption={Unfold to \code{do-while} loop},label={quicksort_inner_loop_do_while},
    language=C]
do {
    ++s;
} while(s < e && *s < *pivot);

do {
    --e;
} while(pivot < e && *e > *pivot);
\end{lstlisting}
\end{minipage}\hfill
\begin{minipage}[t]{.4\textwidth}
\begin{lstlisting}[caption={Rotate to \code{while} loop},label={quicksort_inner_loop_while},
    language=C]
++s;
while(s < e && *s < *pivot) {
    ++s;
}
--e;
while(pivot < e && *e > *pivot) {
    --e;
}
\end{lstlisting}
\end{minipage}

After these changes, our code becomes the listing on the left-hand side below.
We have made slight adjustment so that the ++s and --e statements are gathered together into the
pre-inner-loop section.

\begin{minipage}[t]{.4\textwidth}
\begin{lstlisting}[language=C,caption={After prep steps}]
int* part(int* s, int*e) {
    int* const pivot = s;
    while(true) {
        ++s;
        --e;
        while(s < e && *s < *pivot) {
            ++s;
        }
        while(pivot < e && *e > *pivot) {
            --e;
        }
        if (s < e) {
            swap(s, e); //swap the out-of-place elements
        } else {
            swap(pivot, e);
            return e;
        }
    }
}
\end{lstlisting}
\end{minipage}\hfill
\begin{minipage}[t]{.4\textwidth}
\begin{lstlisting}[language=C,caption={After \emph{pre-inner-loop} relocation}]
int* part(int* s, int*e) {
    int* const pivot = s;
    ++s;
    --e;
    while(true) {
        while(s < e && *s < *pivot) {
            ++s;
        }
        while(pivot < e && *e > *pivot) {
            --e;
        }
        if (s < e) {
            swap(s, e); //swap the out-of-place elements
            ++s;
            --e;
        } else {
            swap(pivot, e);
            return e;
        }
    }
}
\end{lstlisting}
\end{minipage}

After these preparation steps, we are ready to follow the prescribed `loop rotation' procedure.
Namely, relocate the \emph{pre-inner-loop} statements (shown as listing on the right-hand side
above),
convert inner loops into cascading conditionals, wrap up the \emph{post-inner-loop} statements
into an \code{else} clause, and other cosmetic changes (refer to
~\fullref{sec:loop-thinning-technique}\footnote{Note that here
we need to covert the second inner loop to an `\code{else if}' because the first inner loop is
converted to an `\code{if}' clause}).

Shown in ~\autoref{lst:singleLoopHoarePart} is the Hoarse's partition function
after completing the `loop thinning' procedure.
During the refactor process, we relied heavily on the `loop rotation' technique and
the `nested loop thinning' techniques.
We also consciously employed skills for the construction of cascading conditionals.
Comparing with where we started off ~\autoref{lst:nestedLoopHoareMethod},
the new quicksort implementation ~\autoref{lst:singleLoopHoarePart}
retains the optimal and robust runtime as Hoare's algorithm but consists of just one loop as the
Lomuto's partition function does.
That's almost too good to believe.
Our quest for a simple and performant partition scheme finally pays off.

\newpage
\begin{lstlisting}[label={lst:singleLoopHoarePart},language=C,escapechar=\$,
caption={Code after completing the `loop thinning' procedures.}]
int* part(int* s, int*e) {
    int* const pivot = s++;
    --e;$\label{quicksort:pre-increment}$
    while(true) {
         if (s < e && *s < *pivot) {  // find the next element >= pivot
             ++s;
         } else if (pivot < e && *e > *pivot) { // find the prev element <= pivot
             --e;
         } else if (s < e) {
             swap(s++, e--); //swap the out-of-place elements
         } else {
             swap(pivot, e);
             return e;
         }
    }
}
\end{lstlisting}

Additionally, one may use
sentinels to simplify the cluttered conditions in the cascading conditional constructs of
~\autoref{lst:singleLoopHoarePart}.

The end result is listed below.
\begin{lstlisting}[label={lst:hoare-partition-sole-loop-sentinel},
caption={Hoare's partition function with use of sentinel}]
int* part(int* s, int*e) {
    init_swap(s, e);
    int* const pivot = s++;
    --e;
    while(true) {
         if (*s < *pivot) {  // find the next element >= pivot
             ++s;
         } else if (*e > *pivot) { // find the prev element <= pivot
             --e;
         } else if (s < e) {
             swap(s++, e--); //swap the out-of-place elements
         } else {
             swap(pivot, e);
             return e;
         }
    }
}
\end{lstlisting}
Note that this partition function requires at least $3$ elements in the array to work properly.
Interested reader may refer to a more detailed discussion in ~\autoref{sec:quick-sort-sentinel}.

%% file: quicksort_perf_cmp.tex
Note that implementations ~\autoref{lst:singleLoopHoarePart} or
~\autoref{lst:hoare-partition-sole-loop-sentinel} for quicksort is just another way to implement
the quicksort algorithm with Hoare's partitioning scheme.
Their runtime complexity is expected to be the same as that of the traditional ones.
As such, we have designed experiments to test this hypothesis.
To prevent pivot skewness, we use the ``median-of-three'' technique
in all the quicksort implementations use for this experiment\cite{sedgewick1977,sedgewick:1978}.

~\autoref{tab:quicksort-experiment} shows the runtime analysis and comparisons.
The simplified implementation indeed comes with an overhead and is thus slower than
its traditional counterpart of the Hoare's quicksort program.
For sorted data sets (either ascending or descending), there is a $17\%$ slowdown.
But for randomly shuffled data sets, the slow down is consistently around $7\%$.
The experiment data seems to indicate that the simplified
implementation of Hoare's quicksort algorithm shares same runtime complexity as its
traditional counterpart.
Their performance may differ by a multiplier.

\begin{table}[h]
    \centering
    \caption{\label{tab:quicksort-experiment}
    Runtime measurement of Hoare's quicksort algorithms,
    comparing the traditional implementation and the single-loop implementation.
    Experiments are run against three data sets---the first is sorted ascending,
    the second descending, and the third is randomly shuffled.
    Courtesy: The experiment is conducted based on project ``Benchmarking Sorting
    Algorithms''\cite{benchmarking-sorting-algo} with some modification.}
    \vskip 10pt
    \begin{tabular}{r|r|r|r}
        \hline
        \multicolumn{4}{l} {\mbox{Integer arrays in ascending order}}\\
        array size & Traditional & Simplified & Percent difference\\
        \hline
        $10$ & $7$ & $8$ & $14\%$\\
        $100$ & $18$ & $20$ & $11\%$\\
        $1,000$ & $111$ & $126$ & $14\%$\\
        $10,000$ & $1,071$ & $1,239$ & $16\%$\\
        $50,000$ & $4,893$ & $5,867$ & $20\%$\\
        $100,000$ & $9,479$ & $11,131$ & $17\%$\\
        $1,000,000$ & $115,248$ & $137,658$ & $19\%$\\
        \hline
        \multicolumn{4}{l} {\mbox{Integer arrays in descending order}}\\
        array size & Traditional & Simplified & Percent difference\\
        \hline
        $10$ & $7$ & $7$ & $0\%$\\
        $100$ & $18$ & $19$ & $6\%$\\
        $1,000$ & $106$ & $142$ & $34\%$\\
        $10,000$ & $1,020$ & $1,167$ & $14\%$\\
        $50,000$ & $4,757$ & $5,411$ & $14\%$\\
        $100,000$ & $9,378$ & $11,132$ & $19\%$\\
        $1,000,000$ & $118,696$ & $138,311$ & $17\%$\\
        \hline
        \multicolumn{4}{l} {\mbox{Integer arrays randomly shuffled}}\\
        array size & Traditional & Simplified & Percent difference\\
        \hline
        $10$ & $6$ & $6$ & $0\%$\\
        $100$ & $25$ & $25$ & $0\%$\\
        $1,000$ & $264$ & $249$ & $-6\%$\\
        $10,000$ & $2,949$ & $3,065$ & $4\%$\\
        $50,000$ & $16,565$ & $17,943$ & $8\%$\\
        $100,000$ & $34,005$ & $36,400$ & $7\%$\\
        $1,000,000$ & $380,477$ & $406,659$ & $7\%$\\
        \hline
    \end{tabular}
\end{table}

%% file: partition-functions.tex
\section{Hoare's Partitioning Functions}\label{sec:impl-varieties}

Below I compiled a partial list of frequently-made bugs.

\begin{enumerate}
    \item In pivot election, failure to swap pivot element to the beginning of array
    \item For the pointers \code{s} and \code{e}, there are two hesitating options:
    `check-then-increment' or `increment-then-check'.
    For the implementation shown in ~\autoref{lst:nestedLoopHoareMethod},
    first option leads to infinite loop for certain cases.
    \item Also a choice between `\code{<= pivot}' and `\code{< pivot}' for \code{s},
    `\code{>= pivot}' and `\code{> pivot}' for \code{e}.
    Incorrect choice may inadvertently cause pointer incrementation to be skipped under
    obscure circumstances which will, in turn, cause infinite loop
    (Remember that under no circumstances, should either pointer stops approaching each other in
    any outer loop iteration before they meet or cross.)
    \item\label{itm:quicksort-boundary-condition} In the second inner loop for pointer \code{e},
    failing to check boundary condition causes `out-of-boundary' exception.
    For certain solutions, the correct condition is \code{pivot < e}.
    Incorrect boundary condition, such as \code{s < e}, again cause pointer \code{e} to stall
    in the middle of the right partition.
    This will cause the pivot is placed at the wrong place at the return of function.
    \item Outer loop termination logic also needs deliberation.
    The key decision to make is `where to break or return' rather than `when to break or return'.
    For ~\autoref{lst:nestedLoopHoareMethod}, the choice of location of return
    at the end of the loop body was made after quick a few failed attempts.
    \item Fail to swap the pivot element to its final position.
    This step often poses a stumbling block for beginners as well as other unsuspecting
    developers.
    \item Choice of subarray semantics with respect to inclusiveness/exclusiveness of end pointers
    when using start and end pointers to denote a subarray.
\end{enumerate}
Any of the items in this list can easily take a good chunk of debugging time.

A large part of the complication of the problem comes from the fact that this implementation
consists of many moving parts and the design decisions for them are intimately
coupled---changing one would necessitate corresponding changes in others.
For example, suppose we are going to change the subarray semantics of \code{e} from exclusive to
inclusive, we know that the return pointer will have to be different.
But what would it be changed to?
\code{s}, \code{s+1}, \code{e}, or \code{e-1}?
Also, with these changes, the recursive calls in ~\autoref{lst:nestedLoopHoareMethod} line
$4$-$5$ need corresponding changes.
What would that be?
Would it be the LHS or RHS below?

\begin{minipage}[t]{.4\textwidth}
    \begin{lstlisting}[language=C]
        qsort(s, p - 1)
        qsort(p + 1, e)
    \end{lstlisting}
\end{minipage}\hfill
\begin{minipage}[t]{.4\textwidth}
    \begin{lstlisting}[language=C]
        qsort(s, p + 1);
        qsort(p + 2, e);
    \end{lstlisting}
\end{minipage}

Also one may have observed, most of these frequently-made bugs comes with a multiple choice
question.
The collection of them form a tree structure.
Each of the leaf node of the tree structure represents either garbage code or a legitimate
solution.
From an existing solution, one may variational tunnel to nearby solutions.
For more information on how more related implementations may be discovered, please refer to
~\autoref{sec:impl-varieties}.

~\autoref{sec:impl-varieties} where we listed a number of implementations variations
for the partition function.
While not all of them can be simplified into single-loop implementation, we have had success with
a few.
Exactly which one can and which one cannot or why are not completely clear and yet to be
investigated with.

\begin{lstlisting}[label={lst:part_nested_loop_var1},language=C,
caption={Partition scheme for quicksort variation 1.
For elements equal to the pivot, we don't stop '<=' or '<' is not critical.
But '<=' brings more saving on unnecessary swaps.}]
int* part_nested_loop_var1(int* s, int* e) {
    int* const pivot = s;
    while (true) {
        while (s < e && *++s <= *pivot);
        while (pivot < e && *pivot <= *--e);
        if (s < e) {
            swap(s, e);
        } else {
            swap(pivot, e);
            return e;
        }
    }
}
\end{lstlisting}

\begin{lstlisting}[label={lst:part_nested_loop_var2},language=C,
caption={Partition scheme for quicksort variation 2.
Found another variation of partition scheme based on part-nested-loop
Diff wise, it lies between part-nested-loop and part-nested-loop-var
- the order of the two while loops is identical to part-nested-loop
- the boundary checking is identical to part-nested-loop-var
- the return value is shifted toward the left by one unit
}]
int* part_nested_loop_var2(int* s, int* e) {
    int* const pivot = s;
    while (true) {
        while (s < e && *++s <= *pivot);
        while (s < e && *pivot <= *--e);
        if (s < e) {
            swap(s, e);
        } else {
            swap(pivot, s - 1);
            return s - 1;
        }
    }
}
\end{lstlisting}

\begin{lstlisting}[label={lst:part_nested_loop_var3},language=C,
caption={Partition scheme for quicksort variation 3.
Variation of partition scheme.
The two differs in three ways
- order of the two while loops
- increment condition for equal element pointer
- return pointer choice between s and e
}]
int* part_nested_loop_var3(int* s, int* e) {
    int* const pivot = s;
    while (true) {
        while (s < e && *pivot <= *--e);
        while (s < e && *++s <= *pivot);
        if (s < e) {
            swap(s, e);
        } else {
            swap(pivot, s);
            return s;
        }
    }
}
\end{lstlisting}

\begin{lstlisting}[label={lst:part_nested_loop_var4},language=C,
caption={Partition scheme for quicksort variation 4.
Find yet another variation of implementation where the second nested while-loop
is made post-incremental.
This will remove the difficulty encountered in loop-thinning.}]
int* part_nested_loop_var4(int* s, int* e) {
    int* const pivot = s;
    while (true) {
        while (s < e && *pivot <= *--e);
        while (s < e && *s <= *pivot) ++s;
        if (s < e) {
            swap(s, e);
        } else {
            swap(pivot, s);
            return s;
        }
    }
}
\end{lstlisting}

All these listings are to be understood with some pivot-selection mechanism to avoid pivot
skewness.


%% file: use-sentinel.tex

\section{Sentinels in quicksort}\label{sec:quick-sort-sentinel}

The cascading conditional in the loop body of ~\autoref{lst:singleLoopHoarePart}
is still far from straightforward.
A further simplification on that may achievable through the use of
sentinels~\cite{NumericalRecipes:1992}.


For stark effect of the sentinels, we will first fix the traditional implementation of `Hoare's
partition scheme':
~\autoref{lst:QuickSortOutline} gives the outline of the program and
~\autoref{lst:nestedLoopHoareMethod} gives a fully implemented partition function.
As mentioned earlier, one can get involved in the thick of implementing
`Hoare's quicksort algorithm'.
Not only the problem itself is tricky but also the way we implement it.
By picking Hoare's quicksort scheme over its alternative (such as Lomuto's),
we insist on the robust $O(N log N)$ expected runtime.

We attempt to get rid of the boundary checking using sentinels.
These boundary checking are there to ensure that the pointers `\code{s}' and `\code{e}' do not slip
off the ends of the array.
In large arrays, these boundary check operations may get in the way of performance of the algorithm.
But a large part of the reason is to remove clutter from the code.

The main idea is to make boundary checking redundant by effecting some artifacts that catch the
pointers before the `out of boundary' error happens.
This is exactly what sentinel is good at!
By deploying sought-after values, the `sentinels', at the ends of the array before the control
hits the loop, the following will happen:
\begin{enumerate}
    \item the pointers, `\code{s}' and `\code{p}', would have to stop when they hit the ends of
    array;
    \item the subsequent logic will guarantee the termination of the outer loop and
    thus guarantee the inner loop would not be executed again.
\end{enumerate}

What are the sought-after values by `\code{s}' and `\code{e}'?
The out-of-place values!
Particularly, \code{s} is looking for values that are `$\geq$ the pivot';
\code{e} is looking for values `$\leq$ the pivot'.
So then instead of randomly picking one value for pivot, we now randomly pick $3$ values, the median
of which be elected as the pivot as usual, the minimum be deployed to the left end, and the maximum
to the right end.
We group these operations into a function called `\code{init_swap}':
\begin{lstlisting}[]
void init_swap(int* s, int* e) {
    // elect 3 elements
    // sort them as [min, median, max]
    // deploy them at positions: s, (s + 1), and (e - 1)
    ...
}
\end{lstlisting}

Now back the function `\code{part}'.
As we mentioned before, the deployment of sentinels in function `\code{init_swap}' makes
the conditional expressions `\code{s < e}' and `\code{pivot < e}' semantically redundant.
They can now be safely removed.
Below is the function `\code{part}' after all the refactoring is done.
\begin{lstlisting}[]
int* part(int* s, int*e) {
    init_swap(s, e);
    int* const pivot = s;
    while(true) {
        while(*++s < *pivot);   // find the next element >= pivot
        while(*--e > *pivot); // find the prev element <= pivot
        if (s < e) {
            swap(s, e); //swap the out-of-place elements
        } else {
            swap(pivot, e);
            return e;
        }
    }
}
\end{lstlisting}
Comparing with the implementation ~\autoref{lst:nestedLoopHoareMethod}, this code
cleans out clutter in the loop conditions in the inner loops.
We `float' the control all the way through the termination of the program on a `touchless' rail made
by sentinels, without ever needing to check boundary condition.
Of course, the onus is shifted to the initialization before the loop.
Comparing with the inner loops, that section is non-critical.
Not only that the code is made less cluttered, but also the number of such checks
is reduced from $O(N)$ to $O(1)$.
More importantly, by not checking boundary conditions at the most critical section,
we avoid bugs that would otherwise cost us many hours of debugging time.

So by use of sentinels on the Hoare's quicksort algorithm, traditional nested loop implementation
or the single-loop implementation ~\autoref{lst:singleLoopHoarePart} alike,
we shifted the complexity among the nested loops to a non-critical part of the program,
effectively reduced its coding complexity.
Applied to the latter, we arrive at a new level of simplicity for the
implementations of Hoare's quicksort (as shown in ~\autoref{lst:hoare-partition-sole-loop-sentinel}).

%% file: article.bbl
\begin{thebibliography}{10}

\bibitem{Hoare1961}
Charles Antony~Richard Hoare.
\newblock Partition (algorithm 63); quicksort (algorithm 64); find (algorithm
  65).
\newblock {\em Communications of the ACM}, 4:321--322, 1961.

\bibitem{Hoare1962}
Charles Antony~Richard Hoare.
\newblock Quicksort.
\newblock {\em The Computer Journal}, 5:10--16, 1962.

\bibitem{sedgewick:1978}
Robert Sedgewick.
\newblock Implementing quicksort programs.
\newblock {\em Commun. ACM}, 21:847--857, 1978.

\bibitem{Bentley:1986:PP}
Jon Bentley.
\newblock {\em Programming Pearls}.
\newblock ACM, New York, NY, USA, 1986.

\bibitem{top10algorithm:2000}
Jack Dongarra and Francis Sullivan.
\newblock Guest editors introduction: The top 10 algorithms.
\newblock {\em Computing in Science \& Engineering}, 2:22--23, 2000.

\bibitem{data-engineering-avoid-loop}
Antoine Guillot.
\newblock {\em Machine Learning Explained: Vectorization and matrix
  operations}.
\newblock Enhance Data Science, 2018.

\bibitem{Wan:book}
Shoupu Wan.
\newblock {\em Lean Code}.
\newblock TBD, approx. 2019.

\bibitem{sedgewick1977}
Robert Sedgewick.
\newblock The analysis of quicksort programs.
\newblock {\em Acta Informatica}, 7:327--355, 1977.

\bibitem{Bentley-Beautiful-code}
Jon Bentley.
\newblock {\em The most beautiful code I never wrote}.
\newblock O'Reilly Media, 2007.

\bibitem{benchmarking-sorting-algo}
Burak Karakan.
\newblock Benchmarking sorting algorithms.
\newblock https://github.com/karakanb/sorting-benchmark, 2017.

\bibitem{NumericalRecipes:1992}
William~H. Press, Saul~A. Teukolsky, William~T. Vetterling, and Brian~P.
  Flannery.
\newblock {\em Numerical Recipes in C (2Nd Ed.): The Art of Scientific
  Computing}.
\newblock Section ``Quicksort''. Cambridge University Press, 1992.

\end{thebibliography}
